\documentstyle[12pt,aasms4]{article}
\newcommand{\postscript}[2]
 {\setlength{\epsfxsize}{#2\hsize}
  \centerline{\epsfbox{#1}}}
\def\tempest%
{\begin{array}{ccc}
1 & 1 & 1 \\
1 & 1 & 1 \\
4 & 3 & 8
\end{array}}

\begin{document}

\title{Improved Detection Rates for Close Binaries Via Astrometric \\
       Observations of Gravitational Microlensing Events}
\bigskip

\author{Kyongae Chang}
\affil{Department of Physics, \\
       Chongju University, Chongju, Korea 360-764, \\
       kchang@alpha94.chongju.ac.kr,}
\authoremail{kchang@alpha94.chongju.ac.kr}
\bigskip
\centerline{\it \&}
\bigskip

\author{Cheongho Han}
\affil{Department of Astronomy \& Space Science, \\
       Chungbuk National University, Chongju, Korea 361-763 \\
       cheongho@astronomy.chungbuk.ac.kr,}
\authoremail{cheongho@astronomy.chungbuk.ac.kr}

\begin{abstract}
In addition to constructing a Galactic matter mass function 
free from the bias induced by the hydrogen-burning limit, gravitational 
microlensing allows one to construct a mass function which is less 
affected by the problem of unresolved binaries (Gaudi \& Gould).  
However, even with the method of microlensing, the photometric detection of 
binaries is limited to binary systems with relatively large separations 
of $b\gtrsim 0.4$ of their combined Einstein ring radius, and thus the 
mass function is still not totally free from the problem of unresolved 
binaries.  In this paper, we show that by detecting distortions of the 
astrometric ellipse of a microlensing event with high precision 
instruments such as the {\it Space Interferometry Mission}, one can detect 
close binaries at a much higher rate than by the photometric method.
We find that by astrometrically observing microlensing events,
$\sim 50\%$ of binaries with separations of $0.1r_{\rm E}$ can be detected 
with the detection threshold of 3\%.  The proposed astrometric method is 
especially efficient at detecting very close binaries.  With a detection 
threshold of 3\% and a rate of 10\%, one can astrometrically detect binaries 
with separations down to $\sim 0.01r_{\rm E}$.
\end{abstract}

\vskip20mm

\centerline{submitted to {\it The Astrophysical Journal}: Jan 15, 1999}
\centerline{Preprint: CNU-A\&SS-01/99}
\clearpage

\section{Introduction}
   Surveys to discover gravitational microlensing events 
have been and are being conducted toward the Galactic bulge and 
Magellanic Clouds (Alcock et al.\ 1997a, 1997b; Ansari et al.\ 1996; 
Udalski et al.\ 1997; Alard \& Guibert 1997).\markcite{alcock97a, 
alcock97b, ansari96, udalski97}  The experiments are so successful 
that more than $\sim 300$ events have been detected.  As the number 
of events increases, one can construct the mass function of lens matter 
(Han \& Gould 1996; Zhao, Rich, \& Spergel 1996; Gould, 1996; Han \& 
Chang 1998).\markcite{han96, zhao96, gould96, han98}  Since microlensing 
events can occur regardless of the lens brightness, the mass function 
constructed from the result of microlensing experiments is completely 
free from the bias induced by the lens brightness and can be extended 
down to very low mass stars and even brown dwarfs.  By using new 
techniques of pixel method (Ansari et al.\ 1997) and image subtraction 
(Crotts 1992; Tomaney \& Crotts 1996)\markcite{crotts92, tomaney96}, 
the lensing experiments are extended toward unresolved star fields of M31.

   Another important advantage of constructing a mass function with 
microlensing observations is that one can better control the problem 
of unresolved binaries.  If one assumes that the individual masses 
measured are due to single objects without considering unresolved 
binaries, the constructed mass function will be biased toward larger 
masses (Reid 1991; Kroupa, Tout, \& Gilmore 1991).\markcite{reid91,
kroupa91}  With the conventional star count method, it is nearly 
impossible to resolve individual binary components, and thus the 
correction is achieved based on very uncertain models of the fraction 
of binaries and the mass ratios of their components.  On the other hand, 
observations of microlensing events is an efficient method to detect 
binaries.  If an event is caused by a binary lens, the resulting light 
curve differs from that of an event produced by a single-point lens.
Therefore, by detecting the deviations of the light curve from 
the single-lens event light curve, one can detect binaries
(Dominik 1998; Mao \& Paczy\'nski 1991).\markcite{dominik98, mao91}
According to the microlensing detection rate for binaries computed 
by Gaudi \& Gould (1997)\markcite{gaudi97}, a significant fraction of 
binary systems with separations greater than $b\sim 0.4$ of their combined 
Einstein ring radius, $r_{\rm E}$, are detectable with a detection 
threshold of $3\%$.

   However, even with the method of microlensing, the photometric 
detection of binary lenses is limited to binary systems with relatively 
large separations of $b\gtrsim 0.4$.  The typical value of the 
Einstein ring radius for a Galactic bulge event is 
$r_{\rm E}\sim 2\ {\rm AU}$ for a lens mass of $1\ M_{\odot}$.  
According to the model distribution of binary separations determined by 
Duquennoy \& Mayor (1991, see also Figure 1 of Han \& Jeong 
1998)\markcite{duquennoy91, han98}, still a considerable fraction 
($\sim 20\%)$ of binary lenses have separations less than 
$b\sim 0.4r_{\rm E}\sim 1\ {\rm AU}$.  Therefore, although photometric 
observations of binary-lens events provides a superior method 
to conventional methods, the mass function constructed by this method 
cannot be totally free from the problem of unresolved binaries.

   Recently, routine astrometric followup observations of microlensing 
events with high precision instruments such as the {\it Space 
Interferometry Mission} (hereafter SIM, http://sim.jpl.nasa.gov) are 
being discussed as a method to measure the distance and mass of a MACHO 
(Paczy\'nski 1998; Boden, Shao, \& Van Buren 1998).\markcite{paczynski98,
boden98}  When a microlensing event is caused by a single-point lens, 
the observed source star image is split into two, and the location of the 
center of light between the separate images with respect to the source 
star traces out an ellipse (astrometric ellipse; Walker 1995; 
Jeong, Han, \& Park 1999).\markcite{walker95, jeong99}
However, if the lens is composed of binaries, both the number and 
locations of images differ from those of a single-lens event, resulting 
in distorted astrometric ellipse (Safizadeh, Dalal, \& Griest 
1998).\markcite{safizadeh98}

   In this paper, we show that by detecting the distortions of the 
astrometric ellipse, one can detect close binaries at a much higher rate 
than with the photometric method.  We find that by astrometrically 
observing gravitational microlensing events, nearly half of binaries 
with separations of $0.1r_{\rm E}$ can be detected with the detection 
threshold of 3\%.  The proposed astrometric method is especially 
efficient to detect very close binaries.  With the detection threshold 
of 3\% and a rate of 10\%, one can astrometrically detect binaries 
with separations down to $\sim 0.01r_{\rm E}$.

\section{Very Close Binary Lens Events}
   When the lengths are normalized to the combined Einstein ring radius, 
the lens equation in complex notations for a binary lens system is given by 
$$
\zeta = z + {m_{1} \over \bar{z}_{1}-\bar{z}} 
	  + {m_{2} \over \bar{z}_{2}-\bar{z}},
\eqno(2.1)
$$
where $m_1$ and $m_2$ are the mass fractions of individual lenses 
(and thus $m_1+m_2=1$), $z_1$ and $z_2$ are the positions of the lenses, 
$\zeta = \xi +i\eta$ and $z=x+iy$ are the positions of the source and 
images, and $\bar{z}$ denotes the complex conjugate of $z$ (Witt 
1990).\markcite{witt90} The combined Einstein ring radius is related to the 
lens parameters by
$$
r_{\rm E} = \left( {4GM \over c^2}{D_{ol}D_{ls}\over D_{os}}\right)^{1/2},
\eqno(2.2)
$$
where $M$ is the total mass of the binary system, and $D_{ol}$, $D_{ls}$, 
and $D_{os}$ are the distances between observer-lens, lens-source, 
and observer-source, respectively.  The amplification of each image, $A_i$, 
is given by the Jacobian of the transformation (2.1) evaluated at the 
images position, i.e., 
$$
A_i = \left({1\over \vert {\rm det}\ J\vert} \right)_{z=z_i};
\qquad {\rm det}\ J = 1-{\partial\zeta\over\partial\bar{z}}
	                {\overline{\partial\zeta}\over\partial\bar{z}}.
\eqno(2.3)
$$
The images and source positions with infinite amplifications,
i.e.\ ${\rm det}\ J=0$, form closed curves, called critical curves 
and caustics, respectively.  The positions of individual images are obtained 
by numerically solving equation (2.1).  The amplification of a source star 
is given by the sum of the individual amplifications, $A=\sum_i A_i$, 
which are determined by equation (2.3).

   Due to the presence of an additional lens object, the geometry of a 
binary-lens event differs from that of a single-lens event.  In Figure 1, 
we present the geometry of gravitational microlensing for various binary 
lens separations.  In the figure, the two lenses (indicated by two dots 
on the $\xi$-axis) are assumed to have identical masses, i.e.\ $m_1=m_2$, 
and the separation between them in units of the Einstein ring radius, $b$, 
is marked in each panel.  The closed figures drawn with dotted and thick 
solid lines represent  critical curves and caustics, respectively.  
For events with $b=0.1$ and $b=0.4$, there exist two additional triangular 
caustics outside the regions shown.  In addition, there are two small 
circular critical curves near the diamond-shaped central caustic, but 
they are too small to be seen in the figure.  The arclets drawn with a 
thin solid line represent the images of the source star, which is located 
at $(\xi,\eta)=(-0.3,-0.32)$ and is marked by an shaded circle.
The source star is assumed to have a radius of $0.05 r_{\rm E}$.
As the binary separation increases, the critical curve deviates
from a circular Einstein ring and complex caustic structures form.
When the source star is located outside the caustics as shown in the 
figure, there exist three images.  In the figure, the smallest image 
inside the central caustic for small $b$ is too small to be seen.
On the other hand, when a source crosses a caustic, an extra pair of 
images appear or disappear, producing a sharp rise or drop in the light 
curve (Schneider \& Weiss 1986).

   However, if the separation between binary components is very small, 
the geometry of the binary-lens system mimics that of a single-lens system.  
In addition, the light curve imitates that of an event caused by a 
single-point lens with a mass equal to the total mass of the binary and 
located at the center of mass of the binary, making it difficult to 
photometrically distinguish the binary-lens event from a single-lens event.  
The light curve of a single-lens event is represented by
$$
A_0 = {u^2+2 \over u(u^2+4)^{1/2}};\qquad
u = \left[\beta^2 + \left( {t-t_0\over t_E}\right)^2 \right]^{1/2},
\eqno(2.4)
$$
where $u$ is the lens-source separation in units of the angular Einstein 
ring radius $\theta_{\rm E}=r_{\rm E}/D_{ol}$, $\beta$ is the impact 
parameter of the lens-source encounter,
$t_0$ is the time of maximum amplification, and $t_{\rm E}$ is the Einstein 
ring radius crossing time scale.  If the position of the binary center 
of mass is at the origin, $u\equiv \left\vert \zeta\right\vert$.  In the 
lower panel of Figure 2, we present light curves for the binary-lens events 
in the upper left panel of Figure 1, in which the corresponding source star 
trajectories are marked by straight lines.  In the figure, the `x' marks 
represent the amplifications for the binary-lens events, while the smooth 
solid curves are for the corresponding single-lens events.  One finds that 
the light curves can be well approximated by those of single-lens events 
except those with very small impact parameters.  

   On the other hand, the difference in the astrometric shifts of source 
star image centroids ($\vec{\delta\theta}_c$, hereafter centroid shifts) 
between the binary-lens and the corresponding single-lens events is 
considerable compared to the difference in amplifications.  In the upper 
panel of Figure 2, we present the centroid shifts of the binary-lens 
events (marked by `x') for the same source star trajectories in the upper 
panel of Figure 1.  For a single-mass lens event, the centroid shift is 
represented by
$$
\vec{\delta\theta}_{c} = {\theta_{\rm E}\over u^2+2}
\left( u_x \hat{\bf x} + u_y \hat{\bf y} \right),
\eqno(2.5)
$$
where $x$ and $y$ represent the directions parallel and normal to 
the lens-source transverse motion, $u_x=(t-t_0)/t_{\rm E}$ and 
$u_y=\beta$ are the two components of the lens-source separation
vector ${\bf u}$ (Walker 1995).  For comparison of the centroid shifts 
between the binary-lens and corresponding single-lens events, we also present 
the trajectories of the expected centroid shifts of single-lens events, 
i.e.\ astrometric ellipses (smooth solid curves).  For the computation 
of both the amplifications and centroid shifts, we assume a point source.
One finds that the difference in the centroid shifts between the 
binary-lens and single-lens events are considerable even for events 
with large impact parameters.

\section{Detection Rate for Close Binaries}
   In the previous section, we showed by examples that close binaries can be 
more easily detected astrometrically than by the photometric observations 
of microlensing events.  In this section, we statistically compute the binary 
detection rate from the astrometric observations of binary lens events 
and compare it to the rate determined by the photometric method.  For 
direct comparison of the astrometric detection rate, $\Gamma_{\rm ast}$, 
to the photometric detection rate, $\Gamma_{\rm ph}$, computed by Gaudi \& 
Gould (1997)\markcite{gaudi97}, we adopt their method of analysis. 

To analyze how the centroid shift and amplification of a binary-lens 
event deviates from a single-lens event, we define the excess centroid
shift and amplification by
$$
\epsilon_{\delta\theta_c} = 
	{\left\vert\vec{\delta\theta}_{c}-\vec{\delta\theta}_{c,0}\right\vert
	 \over
	 \delta\theta_{c,0} },
\eqno(3.1)
$$
and 
$$
\epsilon_{A} = {A-A_{0}\over A_{0}},
\eqno(3.2)
$$
where $\vec{\delta\theta}_{c,0}$ is the centroid shift for a 
single-lens event.  Since the centroid shift is a two-dimensional vector, 
we compute the excess centroid shifts based on the absolute value of the 
difference between $\vec{\delta\theta}_{c}$ and $\vec{\delta\theta}_{c,0}$, 
not on the sutraction of their scalar values.  We then compute the values 
of $\epsilon_{\delta\theta_c}$ and $\epsilon_{A}$ as a function of source 
position, i.e.\ $(\xi,\eta)$.  For the computation of 
$\epsilon_{\delta\theta_c}$ and $\epsilon_{A}$, we assume a point source.  
For the discussion about the finite-size source effect, see \S\ 6 of Gaudi 
\& Gould (1997).\markcite{gaudi97}  In Figure 3, we present the contours 
of excess centroid shift for events with various binary separations and 
mass ratios $q=m_{1}/m_{2}$ between the binary components.  In the figure, 
contours are drawn at the levels of $\epsilon_{\delta\theta_{c}}= 0.30$ 
(thickest lines), $0.10$, and $0.03$ (lightest lines).  
The contours of excess amplification for events with the same binary 
separations and mass ratios are found in Figure 1 of Gaudi \& Gould 
(1997)\markcite{gaudi97}, in which the contours are 
drawn at the levels of $\epsilon_{A}=\pm 0.10$ and $\pm 0.03$.   
One finds that the region of significant centroid shift deviation is much 
larger than the region of amplification deviation measured at the same 
level of excess.  In Figure 4, we also present contours of excess centroid 
shifts for binaries with very small separation ($b\leq 0.1r_{\rm E}$).
From the figure, one finds that the regions of noticeable excess centroid
shift is considerable even for very small values of $b$.

   Once the contours of excess amplification and centroid shift are 
constructed, the binary detection rates are obtained by computing the 
ratio of the number of events whose source locations are within the 
excess region during the measurements to the total number of trial events.  
The source trajectory orientations with respect to the projected binary 
axis, $\theta$, and the impact parameters, $\beta$, are randomly chosen in 
the ranges of $0\leq \theta\leq 2\pi$ and $0\leq \beta\leq 1$.  Measurements 
for each event is assumed to be performed 25 times during the event, which
corresponds to one measurement per day for a Galactic bulge event caused by 
a total mass of $M\sim 0.4\ M_\odot$ (Gaudi \& Gould 1997).\markcite{gaudi97}
We consider a binary to be detected if the excess is greater than some
threshold excess values: $\epsilon_{A}$ and $\epsilon_{\delta\theta_c}$
for photometric and astrometric measurements respectively.
Following Gaudi \& Gould (1990)\markcite{gaudi97}, we adopt the two 
threshold values of 0.03 and 0.1.

   Figure 4 shows the binary detection rates as a function of binary 
separation and mass ratio expected from the photometric and astrometric 
observations of microlensing events.  To better show the detection rates 
for binaries with very small separations and mass ratios, we present 
the values of $b$ and $q$ on a logarithmic scale.  Contours have equal spacing 
of 10\% and the threshold excess amplification and centroid shift are 
marked in each panel.  From the comparison of detection rates between 
the two methods, one finds that the astrometric detection rates are 
significantly larger than the photometric rates in all regions of $b$--$q$
space.  For example, while the photometric detection rates for a binary 
lens with $(b,q)=(0.3,0.1)$ are $\Gamma_{\rm ph}< 10\%$ and $\sim 15\%$ 
with the threshold excesses of $\epsilon_A = 0.1$ and 0.03 respectively, 
the astrometric detection rates for the same binary are 
$\Gamma_{\rm ast}\sim 70\%$ and $\sim 95\%$ with the same thresholds.
Our proposed method is especially efficient at detecting very close binaries.
With a detection threshold of $\epsilon_{\rm as}=0.03$ and a rate of 10\%, 
one can astrometrically detect binaries with separations down to 
$\sim 0.01r_{\rm E}$.

\section{Summary}
   Gravitational microlensing provides us with an important method of 
constructing the Galactic matter mass function which is free from the biases 
induced by the hydrogen burning limit.  In addition, by detecting 
the deviations in light curves due to binary systems, one can construct 
a mass function that is less affected by the problem of unresolved binaries.  
However, since the light curves of binary-lens events with very small 
separations mimic those of single-lens events, the photometric 
detection of binaries is limited to systems with relatively large binary 
separations.  Therefore, although photometric microlensing observations  
provide a superior method of detecting binaries, the constructed mass 
function will not be totally free from the problem of unresolved binaries.
However, by astrometrically measuring the distortions of astrometric 
ellipses, one can not only significantly improve the binary detection rate 
but also detect binaries with much smaller separations.  Therefore, 
astrometric observations of microlensing events can significantly reduce the 
uncertainties in the mass function caused by unresolved binaries.

\acknowledgements
We would like to thank to P.\ Martini and M.\ Everett for careful reading 
the manuscript.  We also would like to thank B.\ S.\ Gaudi \& A.\ Gould 
for kindly providing a program to compute amplifications of binary-lens 
events.  Chang, K.\ was supported by a Korean Science and Engineering 
Foundation (KOSEF) grant 971-0203-012-2.

\clearpage

\noindent
{\footnotesize {\bf Figure 1:}\ 
The geometry of gravitational microlensing for various binary lens 
separations.  In the figure, the two lenses with equal masses are 
located on the $\xi$-axis (indicated by two dots) and the separation 
$b$ between them is marked in each panel.  
The closed figures drawn with dotted and thick
solid lines represent  critical curves and caustics, respectively.
For events with $b=0.1$ and $b=0.4$, there exist two additional triangular
caustics outside the regions shown.  In addition, there are two small
circular critical curves near the diamond-shaped central caustic, but
they are too small to be seen in the figure.  The arclets drawn with a
thin solid line represent the images of the source star, which is located
at $(\xi,\eta)=(-0.3,-0.32)$ and is marked by an shaded circle.
The source star is assumed to have a radius of $0.05 r_{\rm E}$.
When the source star is located outside the caustics as shown in the
figure, there exist three images.  However, the smallest image
inside the central caustic for small $b$ is too small to be seen.
The straight lines in the panel for $b=0.1$ represent the source 
star trajectories which are responsible for the astrometric shifts and
light curves in Figure 2.
} \clearpage

\postscript{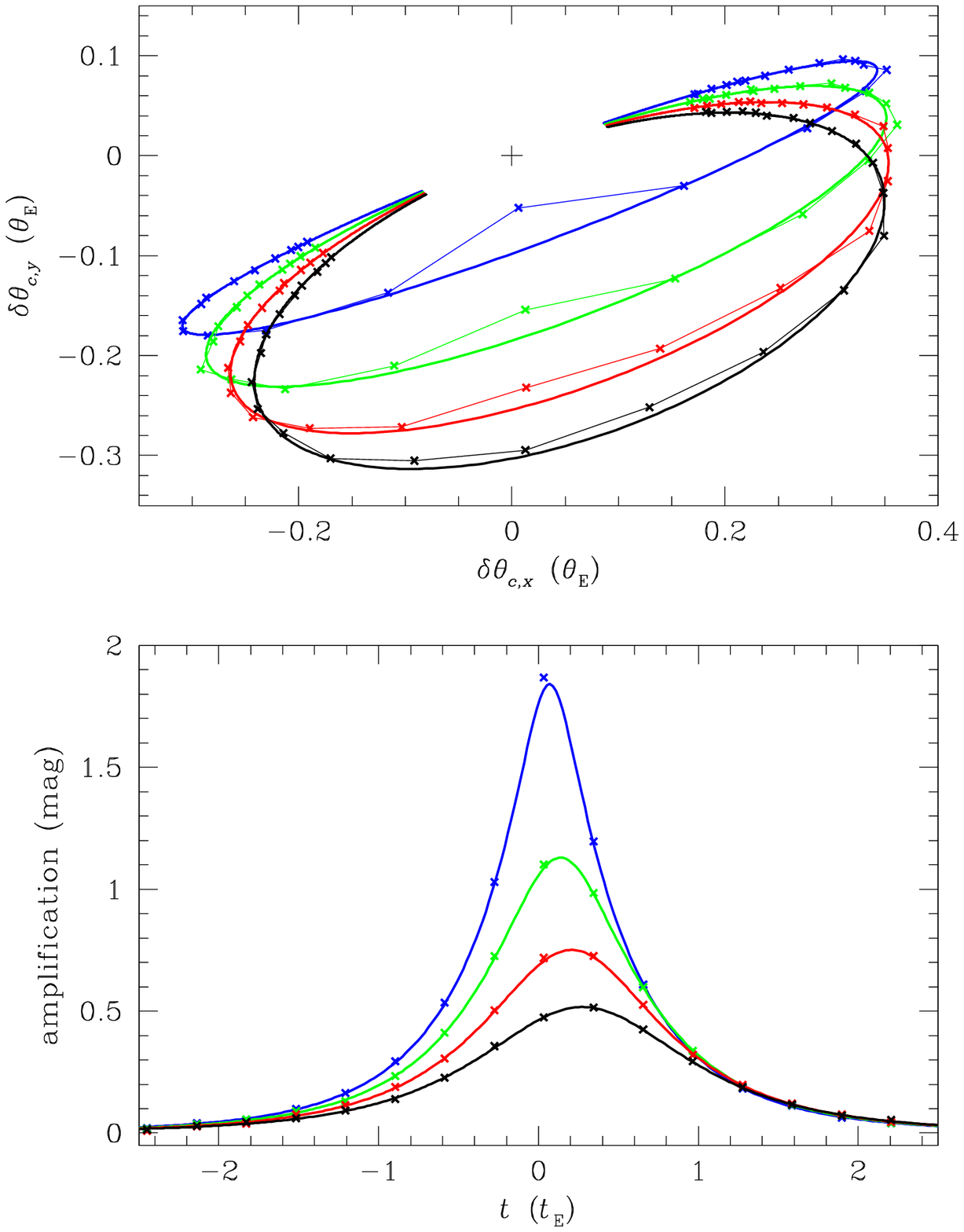}{1.0}
\noindent
{\footnotesize {\bf Figure 2:}\ 
The astrometric light centroid shifts and light curves of close 
binary-lens events.  The centroid shifts and light curves are 
for the binary-lens event in the upper left panel of Figure 1, in which 
the corresponding source star trajectories are marked by straight lines.  
The `x' symbols mark the centroid shifts and amplifications for the 
binary-lens event, while the smooth solid curves are for a single-lens 
event with a mass equal to the total mass of the binary and located 
at the center of mass of the binary.
} \clearpage

\postscript{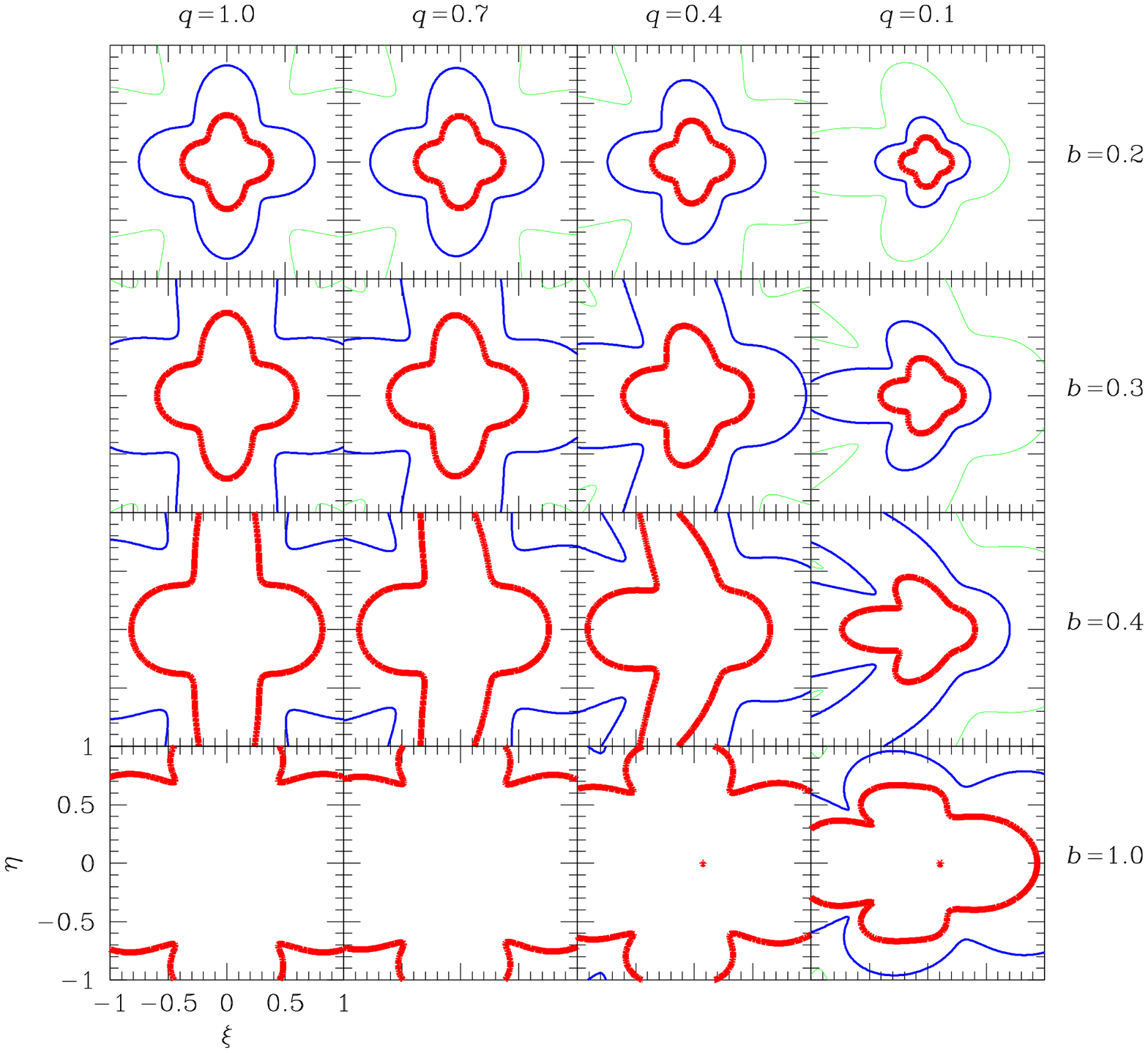}{1.0}
\noindent
{\footnotesize {\bf Figure 3:}\ 
Contours of excess centroid shift, $\epsilon_{\delta\theta_{c}}$,
for events with various binary separations and the mass ratios between
the binary components.  Contours are drawn at the levels of 
$\epsilon_{\delta\theta_{c}}=0.30$ (thickest lines), 0.10,
and 0.03 (lightest lines).  The positions of the masses are 
chosen so that the center of mass is at the origin, but both masses are 
on the $\xi$-axis, and the smaller mass is to the left.  For direct 
comparison with the contours of excess amplification of Gaudi \& Gould 
(1997), we choose the same positions of masses, binary separations, 
and binary-component mass ratios.
} \clearpage

\postscript{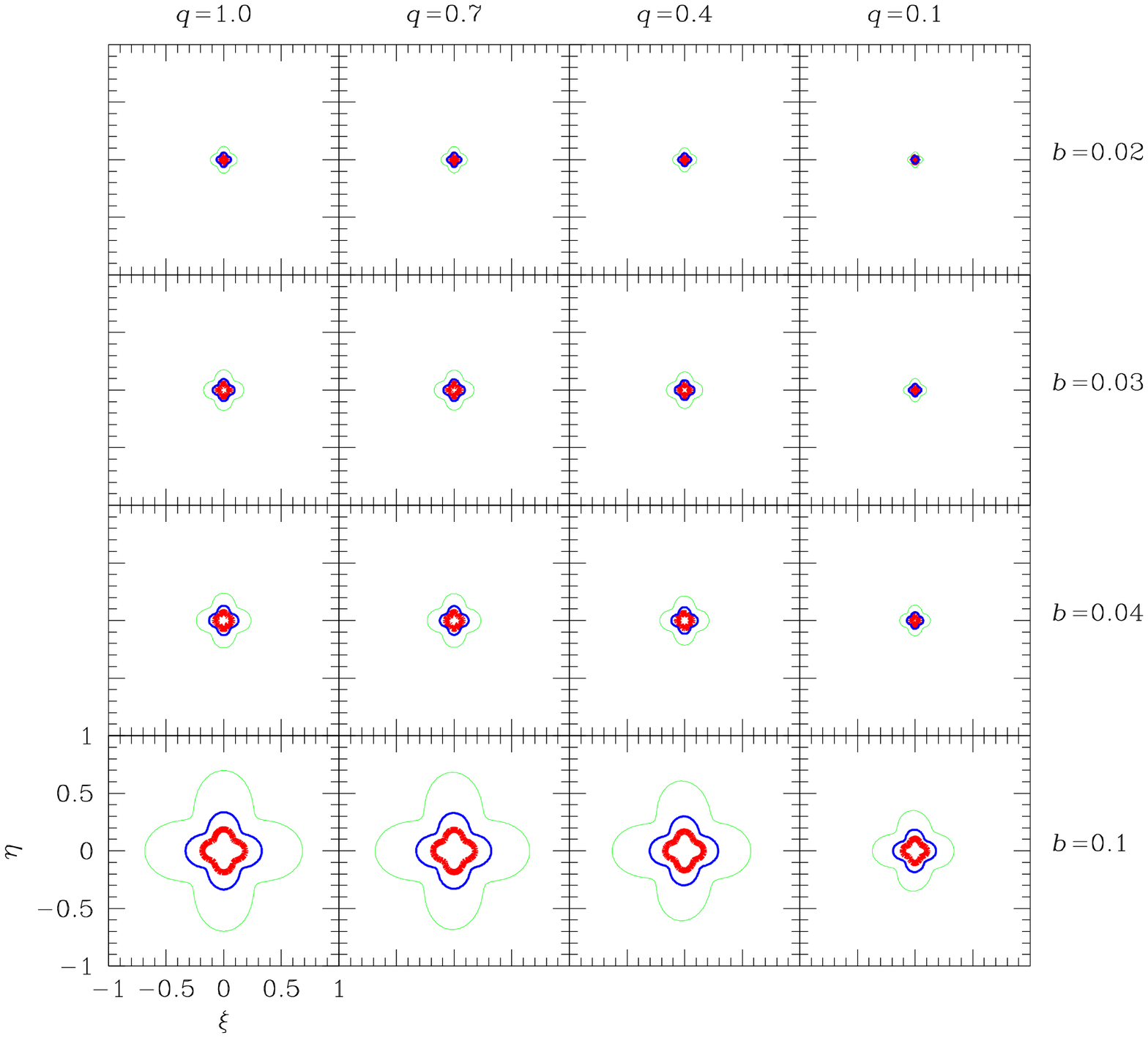}{1.0}
\noindent
{\footnotesize {\bf Figure 4:}\ 
Contours of excess centroid shift, $\epsilon_{\delta\theta_{c}}$,
for binary-lens events with very close binary separations.
The positions of the masses and the levels of the contours are
chosen in the same way as in Figure 3.
} \clearpage

\postscript{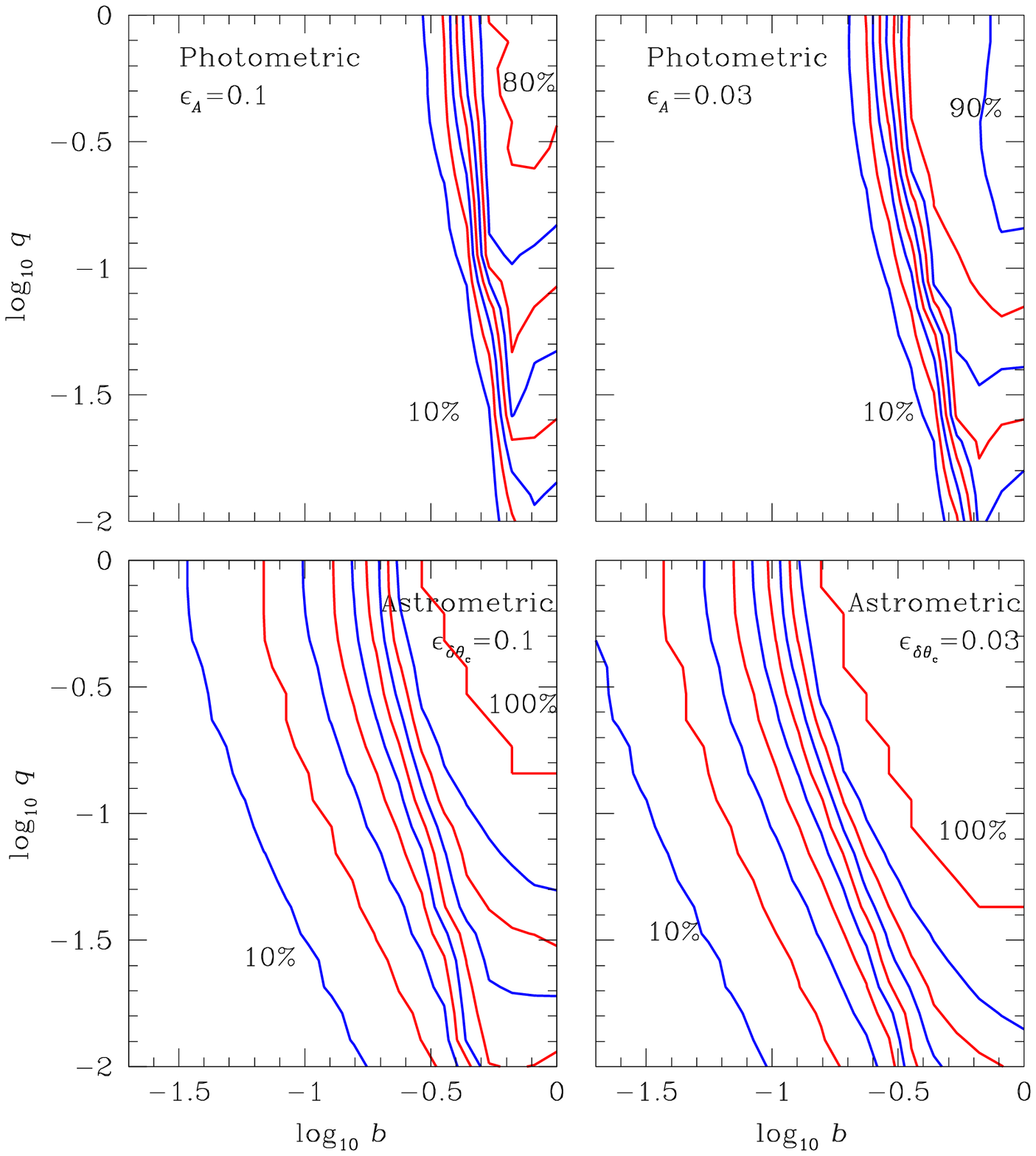}{1.0}
\noindent
{\footnotesize {\bf Figure 5:}\ 
Binary detection rates as a function of binary separation and mass ratio 
expected from the photometric and astrometric observations of microlensing 
events.  Contours have equal spacing of 10\%.  To better show the event rates 
for binaries with very small separations and mass ratios, the values of 
$b$ and $q$ are presented in logarithmic scale.  A binary is considered 
to be detected if the excess of amplification (and the centroid shift)
is greater than the threshold value (marked in each panel) 
during the measurements.
}\clearpage

\end{document}